\documentclass{article}
\usepackage{spconf,amsmath,graphicx}
\usepackage{epsfig} 
\usepackage{mathptmx} 
\usepackage{times} 
\usepackage{amssymb}  
\usepackage{cleveref}

\title{A NOVEL FORWARD-PDE APPROACH AS AN ALTERNATIVE TO EMPIRICAL MODE DECOMPOSITION}
\name{Heming Wang$^{\dagger}$, Richard Mann$^{\star}$ and Edward R. Vrscay${^\dagger}$}
\address{$^{\dagger}$Department of Applied Mathematics, \/ $^{\star}$David R.~Cheriton School of Computer Science \\ University of Waterloo, Ontario, Canada N2L3G1}
\begin{document}
%
\maketitle
\begin{abstract}

In this paper we present a mathematical model of the Empirical Mode Decomposition (EMD). Although EMD is a powerful tool for signal processing, the algorithm itself lacks an appropriate theoretical basis. The interpolation and iteration processes involved in the EMD method have been obstacles for mathematical modelling. Here, we propose a novel forward heat equation approach to represent the mean envelope and sifting process. This new model can provide a better mathematical analysis of classical EMD as well as identifying its limitations. Our approach achieves a better performance for a ``mode-mixing'' signal as compared to the classical EMD approach and is more robust to noise. Furthermore, we discuss the ability of EMD to separate signals and possible improvements by adjusting parameters.

\end{abstract}


\section{Introduction}
\label{sec:intro}

Huang et al. \cite{Huang-EMD-98} introduced the empirical mode decomposition (EMD) in 1998 as a tool to analyze linear and non-stationary signals. EMD has been applied quite successful in science and engineering.  It treats a signal as a mixture of mono-components and applies a \textit{sifting process} to separate different modes of oscillation which are referred to as Intrinsic Mode Functions (IMF). EMD is essentially a decomposition algorithm that extracts the highest local frequency components from the signal for each IMF. A repeated application produces a decomposition of a signal into of components with decreasing frequency. The Hilbert transform is then applied to each component in order to determine instantaneous frequencies. The amplitudes and instantaneous frequencies may then be combined to produce a local time-frequency analysis of the signal.  

The ability of EMD to capture intrinsic physical features for non-stationary signals \cite{Variant-EMD} has been demonstrated for real-world signals with limited frequency components, such as signals obtained from earthquakes \cite{Huang-EMD-98}, medical experiments \cite{EEG} and rotating machinery \cite{Rotating-Machine}. There are some kinds of signals, however, for which the sifting process fails to separate into different oscillatory modes. Because most of the work on EMD has focused on algorithms as opposed to mathematical analysis, there has been very little work on developing a rigorous theoretical basis for EMD as well as an understanding of why it fails for certain kinds of signals.  The need for a mathematical model  which explains the principle of EMD and provides a description of the region where it can work effectively has been the motivation for this work.

One major obstacle for mathematical modelling of EMD is the interpolation process employed by the algorithm. In perhaps the first effort to model the EMD interpolation procedure \cite{El-pde}, a rather large number of variables were encountered. This, plus the fact that iteration is involved, makes it difficult to arrive at an accurate expression in the model.  In this paper, we
propose a forward heat PDE approach to solve these problems. Instead of taking the average of two envelopes, a forward heat equation is introduced to construct a mean envelope. 
This mean envelope can be viewed as the result of passing a signal through a smooth filter - in this case, a Gaussian filter. After obtaining the mean envelope,  we repeat the same steps as those of the original EMD sifting process in order to extract the IMFs.  Our approach generates results which are similar to classical EMD, but provides a solid mathematical basis for the method.  

The remainder of the paper is organized as follows. In Section 2, we briefly review the details of classical EMD method and some related work. In Section 3, we introduce the new forward-heat PDE approach. In Section 4, mathematical interpretation of EMD is provided, and the limitations are analyzed. In Section 5, numerical implementation and experimental results are presented.

\section{RELATED WORKS}

\subsection{CLASSICAL EMD ALGORITHM}

The classical EMD algorithm may be summarized as follows:\\[5pt]
{\bf 1.}  Find all local maximal and minimal points of the signal $S(x)$. \\[5pt]
{\bf 2.}  Interpolate between maximal points to obtain upper envelope function 
$E_{upper}(x)$ and between minimal points to obtain lower envelope function $E_{lower}$(x). \\[5pt]
{\bf 3.}  Computer the local mean: $m(x) = \frac{1}{2}(E_{upper}(x)+E_{lower}(x))$. \\[5pt]
{\bf 4.}  Extract mean from signal:  $c(x) = S(x) - m(x)$. \\[5pt]
{\bf 5.}  If $c(x)$ is not an IMF, iterate 3) and 4) until it is. \\[5pt]
{\bf 6.}  After finding IMF, subtract it from $S(x)$ and repeat Step 2 to
obtain the residual.\\[10pt]
\noindent
There are, however, several drawbacks \cite{huang2005introduction}:\\[5pt]
\noindent
{\bf Vague Definition of IMF}
which presents obstacles in implementation. For an IMF, the number of extrema and zero-crossings must differ at most by one. In addition, the ``local mean'' of the IMF should be close to zero.  It is therefore necessary to choose appropriate stopping criteria for the sifting process.\\[5pt]
\noindent
{\bf Boundary Effects:} 
Proper boundary conditions are necessary in order to minimize errors at the
		boundaries.  Otherwise, there can be ``tweaking'' at the
		endpoints.\\[5pt]
\noindent
{\bf Mode Mixing:}
Whenever the signal contains riding waves, some frequency components will vanish after performing EMD. In an effort to solve this problem, Huang introduced a new method called ensemble empirical mode decomposition \cite{EEMD}, in which Gaussian noise is first added, and the signal then denoised.

\subsection{Backward Heat Equation}
As mentioned earlier, interpolation represents an obstacle in the mathematical modelling of EMD \cite{el2013pde}. A PDE approach was proposed in \cite{El-pde, el2009pde} to overcome
this obstacle.  Here, for a prescribed $\delta > 0$, 
the upper and lower envelopes of a function 
$h(x)$ are defined as
follows,
\begin{equation}
	U_\delta (x) = \sup_{|y|<\delta} h(x+y) \, , 
	~~~ L_\delta (x) = \inf_{|y|<\delta} h(x+y) \, .
\end{equation}
After 
Taylor expansions are applied to the envelopes, the mean envelope 
is defined as
\begin{equation}
m_\delta(x) = \frac{1}{2}(U_\delta (x)+L_\delta (x)) \approx h(x) + \frac{\delta^2}{2} h^{''}(x) \, .
\end{equation}
The sifting process -- the process to obtain the 
Intrinsic Mode Function (IMF) -- is then defined as follows,
\begin{equation}
h_{n+1}(x) = h_{n}(x) - m_\delta(x) \, , ~~~ h_0 (x) = S(x) \, .
\end{equation}
Using the following Taylor expansion in $t$,
\begin{equation}
h_{n+1} = h(x,t+ \Delta t) = h_n + \Delta t \frac{\partial h}{\partial t} + O(\Delta t^2) \, ,
\end{equation}
the authors arrive at the following PDE,
\begin{equation}
\label{backwardpde}
\frac{\partial h}{\partial t} + \frac{1}{\delta^2} h + \frac{1}{2}\frac{\partial^2 h}{\partial x^2} = 0  \, , ~~~ ~~ 
h(x,0) = S(x) \, ,
\end{equation}
which is known as a {\em backward heat equation} since
the diffusivity constant is negative. (Note that the initial condition, $h(x,0)$, to this PDE is
the original signal $S(x)$.) Unfortunately, there are several drawbacks to this approach:\\[5pt]
\noindent
{\bf 1.}  The parameter $\delta$, which is chosen empirically, has a significant influence on the result. For a generalized signal $s = \sum_k A_k \cos(\omega_k x + \phi_k) $, the solution is
\begin{equation}
h(x,T) = \sum_k e^{(\frac{\omega_k^2}{2} - \frac{1}{\delta^2})T} A_k \cos(\omega_k x + \phi_k) \, .
\end{equation}
As $T$ increases,
		the amplitudes of components with lower frequencies 
$\omega < \frac{\sqrt{2}}{\delta}$ will be decreased 
		at each step and therefore vanish at the
		end of the algorithm. Only the higher frequencies
		$\omega \geq \frac{\sqrt{2}}{\delta}$ survive.
		Therefore, choosing $\delta$ 
		requires an additional knowledge of the signal.\\[5pt]
\noindent
{\bf 2.}
Even if we extract correct the frequency component from the signal, we cannot guarantee that the amplitude of the component is correct. In order to distinguish two frequency components, one sometimes has to decrease their amplitudes to very small values.

\noindent
{\bf 3.}
As mentioned earlier, Eq. ~(\ref{backwardpde}) is a backward heat/diffusion equation.  Because the diffusivity parameter is negative, the evolution of a signal will be {\em opposite} to that of a signal under the standard (forward) diffusion PDE -- signals become less smooth and local amplitudes grow exponentially.  As expected, numerical methods also suffer from instability.

\section{FORWARD PDE ALGORITHM}
\label{sec:forwardPDE}	

\subsection{Introduction}
\begin{figure}[h]
\centering
\includegraphics[width=0.38\textwidth]{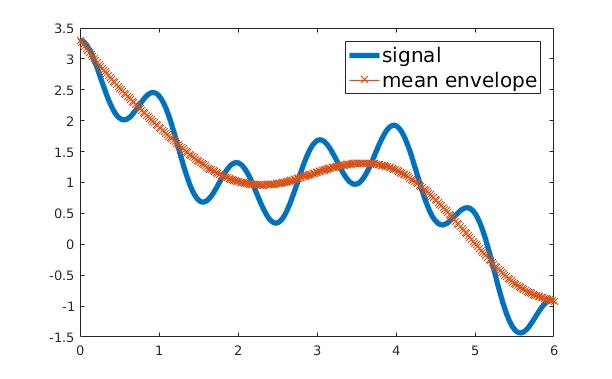}
\caption{Mean Envelope Obtained by Forward Heat Equation}
\label{Fig:mean_envelope_demo}
\end{figure}

We now outline our PDE-based method to perform a new type of interpolation in the EMD algorithm. The idea is very simple:
Instead of taking the average of two envelope functions of a signal $S(x)$
to produce a mean (Step 3 in Section 2.1),
we proceed as follows.
For prescribed values of the diffusivity constant $a > 0$
and time $T > 0$ (which can be adjusted, as will be discussed below), 
solve the following initial value problem
for the heat/diffusion equation,
\begin{equation}
	\label{heateqn}
	\frac{\partial h}{\partial t} = a \frac{\partial^2 h}{\partial x^2} \, , ~~~~
h(x,0) = S(x) \, 
\end{equation}
and define the {\em mean curve} of $S(x)$ to be $m(x) = h(x,T)$.
(Of course, $m(x)$ is equivalent to the convolution of $S(x)$ with the
Gaussian function with standard deviation $a$.)
One of the primary motivations for this definition is that the
time rate of change of $h(x,t)$ is zero at spatial inflection points of $h$.
An example is shown in Figure \ref{Fig:mean_envelope_demo}.
This is the basis of the following modified EMD algorithm applied
to a signal $S(x)$:\\[5pt]
\noindent
{\bf 1.}  Initialize:  Let $n=0$ and set $h_0(x,0) = S(x)$.\\[5pt]
{\bf 2.}  Find mean of $h_n(x,0)$:  
Solve the PDE in (\ref{heateqn}) for $h_n(x,t)$ for 
$0 \leq t \leq T$.  Then define $m_n(x) = h_n(x,T)$.\\[5pt]
{\bf 3.}  Extract mean:  
Define $c_n(x) = h_n(x,0) - h_n(x,T)$.\\[5pt]
{\bf 4.} If $c_n$ is not an IMF,  
let $h_{n+1}(x,0)=c_n(x)$, $n \to n+1$ and
go to Step 2.\\[5pt]

\subsection{Parameter selection}
\subsubsection{How to choose parameter a}
Parameter $a$ is crucial to determine the mean envelope so it must
be carefully chosen. To ensure that the mean envelope is always within the range of the signal amplitude, it is necessary that $a \leq \frac{1}{\omega^2}$.  
As mentioned in \cite{EMD-sampling}, if the sampling rate $f_s$ is not sufficiently large, sampling effects will cause a loss of accuracy. We must
assume that $f_{max}$, the maximum frequency to
be extracted, satisfies $f_{max} < f_s$.  This
implies that $\frac{1}{4\pi^2 f_s^2} < \frac{1}{\omega_{max}^2}$.
It is then safe to set $a = \frac{1}{4\pi^2 f_s^2}$. Ideally the parameter $a$ should be set to $a = \frac{1}{\omega_{max}^2}$. There are two practical approach to estimate $a$:  (1) autocorrelation, (2) zero-crossing rate. 

\subsubsection{The pair of parameters T and N}
The parameters $P$ and $T$ determine the shape of the mean curve, and $N$ represents the number of iterations. In order to be able to separate the high-frequency component from the other components, we
impose the following condition,
\begin{equation}
	\left [ \frac{1-e^{-f_0^2T}}{1-e^{-T}} \right ]^N  = \delta  \, ,
\end{equation}
where $\delta > 0$, an adjustable parameter, is close to zero.
Here, $f_0$ is the {\em cutoff frequency ratio}:  If we let $0 < f < 1$ denote
the ratio between two frequency components (lower/higher), then the
algorithm may fail to separate the components $f > f_0$, i.e.,
if the two components are too close to each other.
When $f < f_0$, the ratio of the norms of the lower- and higher-frequency
components will satisfy $||\frac{S_{lower}}{S_{higher}}|| < \delta$. 
Therefore, finding the optimal parameters reduces to the following 
two problems,

\begin{equation}
	f_0 = \left [ \frac{\log(1-(\frac{\delta}{\alpha})^{\frac{1}{N}} (1-\epsilon) )}{\log \epsilon} \right ]^{1/2} ~ , ~~
(1-e^{-T})^N = 1-\delta \, .
\end{equation}

Under these restrictions, we seek to maximize $f_0$. Assume that $a = \frac{1}{\omega_{max}}$, and let $e^{-T} = \epsilon$. Also assume that the two signals are sufficiently separated, i.e., $||\frac{S_{lower}}{S_{higher}}|| < \delta$.

\begin{figure}[thpb]
\centering
\includegraphics[width=0.38 \textwidth]{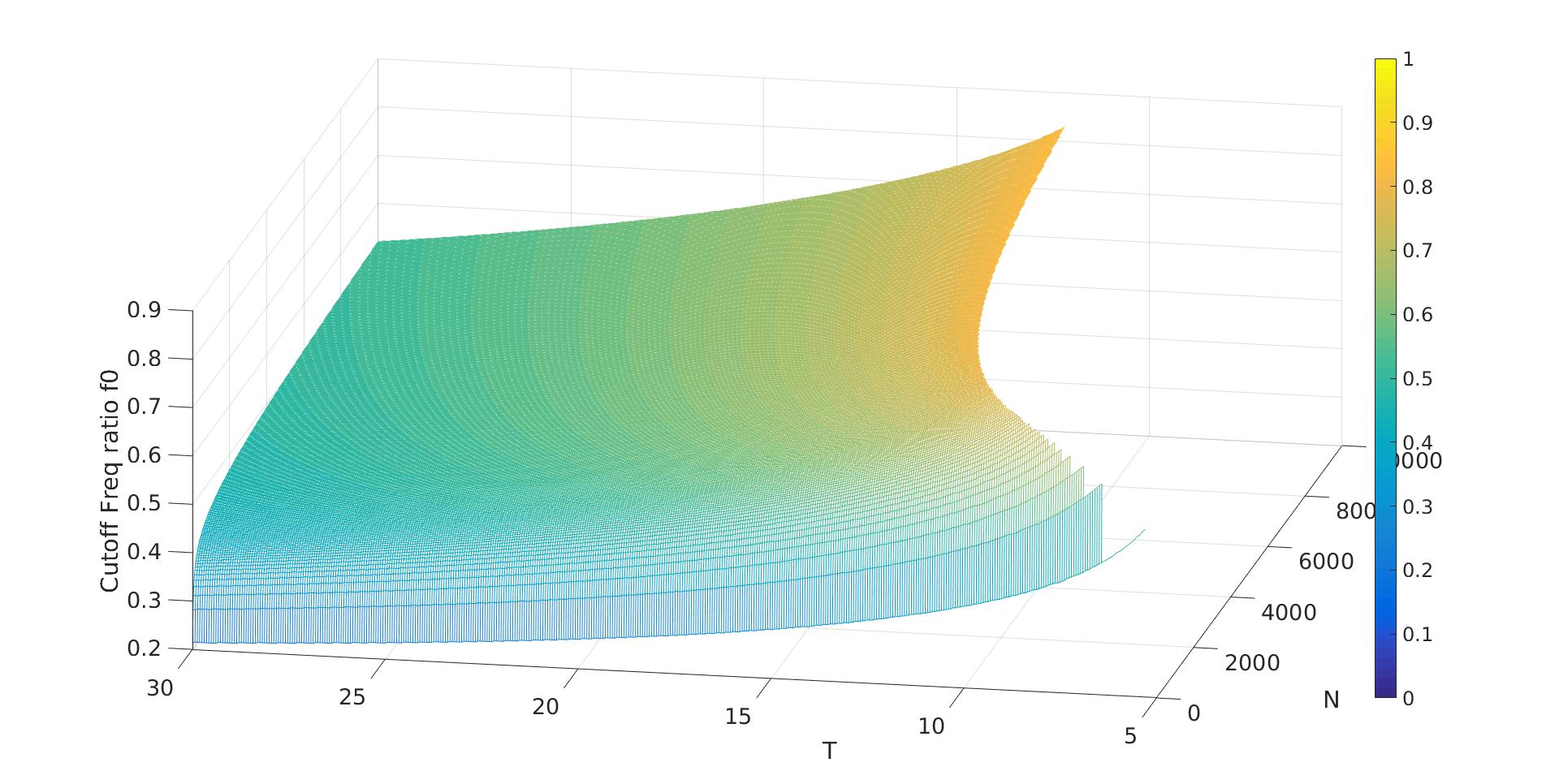}
\caption{Cutoff frequency varied with parameter $T$ and $N$}
\label{Fig:params-study}
\end{figure}

The results of one numerical experiment, with 
$N = 100.0$ and $T = 10.0$, are shown in Figure \ref{Fig:params-study}.
From these results, we conclude that the 
theoretical cutoff frequency should be $f_0 \simeq 0.7$.

\section{Mathematical explanation of EMD and its limitations} 

\subsection{Forward PDE interpretation of EMD}
Here we consider the following simple model which is sufficiently
general to represent many realistic signals,
\begin{equation}
	S(x) = \sum_{k=1}^K  A_k{cos(\omega_k x + \phi_k)} + C =
	\sum_{k=1}^K  s_k(x) + C \, .
\end{equation}
Solving Eq.~(\ref{heateqn}) for first mean envelope PDE, we obtain

\begin{equation}
	m_a(x,T) = \sum_{k=1}^K e^{-a\omega_k^2 T}s_k + C
\end{equation}
After $N$ iterations,
our modified EMD algorithm yields the
following result for the $k$th cosine component $s_k(x)$, 

\begin{equation}
h_{k,N} = (1-e^{-a\omega_k^2T}) h_{k,N-1} = (1-e^{-a\omega_k^2T})^{N} s_k  \, .
\end{equation}

Now suppose, without loss of generality, that
$\omega_1 < \omega_2 < \cdots < \omega_K = \omega_{max}$.
It is easy to show that
for $N$ sufficiently large,
\begin{equation}
	h_N = \sum_{k=1}^N (1-e^{-a\omega_k^2T})^N s_k 
	\simeq (1 - e^{-a \omega_K^2T} )^N s_K  \simeq s_K \, ,
\end{equation}
where the final approximation is valid for $T$ sufficiently large.
By choosing the appropriate set of parameters, the IMF extracted 
after $N$ iterations will be (at least approximately) 
the highest-frequency component $s_K$.

\subsection{Border effect}
Border effects arise mostly at the mean curve procedure , which involves the solution of the heat PDE (Eq.~(\ref{backwardpde})). Unlike the classical approach, we can control the boundary effect by imposing appropriate boundary condition on the PDE. For example, by assuming the signal is periodic, or fixed at both ends. 

\subsection{View of Filter}
As stated by Fladrin \cite{Flandrin-filter}, the EMD algorithm is equivalent to a set of filter banks, which is justified in our PDE method. In each iteration of our PDE approach, the mean of the signal is obtained by passing it through a low-pass filter. 
Subsequent subtraction of the mean from the signal is therefore
equivalent to passing it through a high-pass filter.


\section{Numerical Results}

\subsection{Two-mode mixing}
\begin{figure}[t]
\begin{tabular}{ll}
\includegraphics[width = 0.23\textwidth]{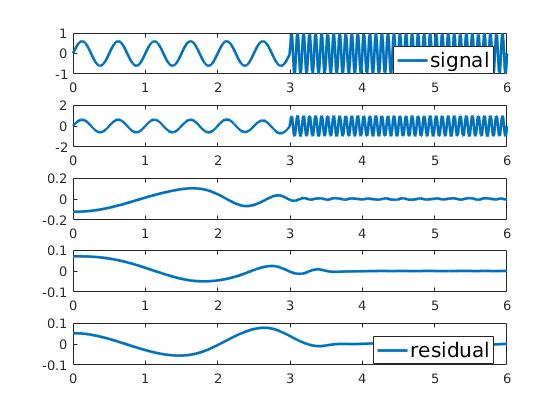}
&
\includegraphics[width = 0.23\textwidth]{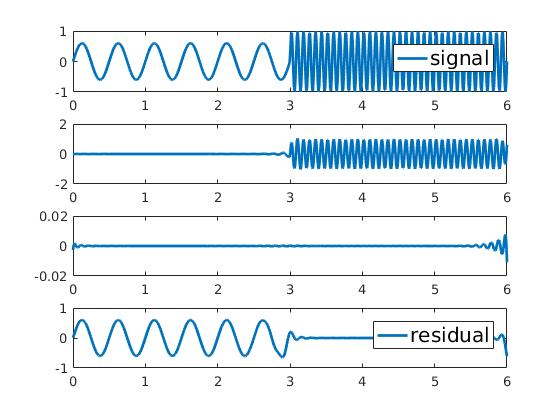}
\end{tabular}
\caption{Experiment on mode-mixing signal by classical EMD (left) and forward-PDE approach (right)}
\label{Fig:mode-mixing-signal}

\end{figure}

\begin{figure}[h]
\begin{tabular}{ll}
\includegraphics[width = 0.23\textwidth]{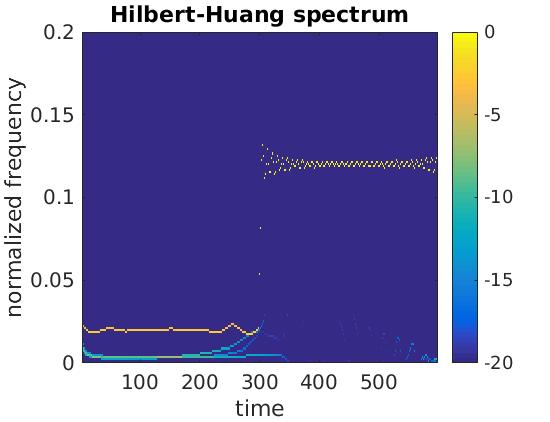}
&
\includegraphics[width = 0.23\textwidth]{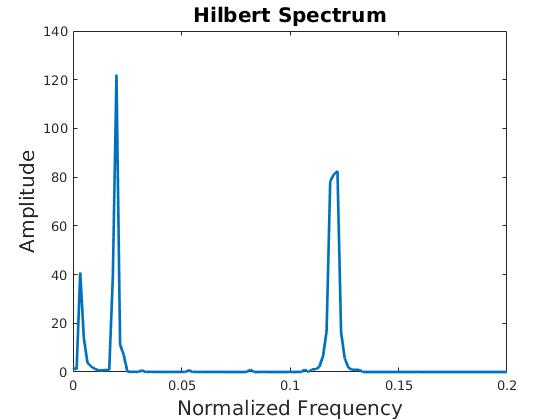}
\end{tabular}
\caption{Hilbert-Huang Spectrum for mode-mixing signal using classical EMD approach. }
\label{Fig:emd-mode-mixing}

\end{figure}

\begin{figure}[h]
\begin{tabular}{ll}
\includegraphics[width = 0.23\textwidth]{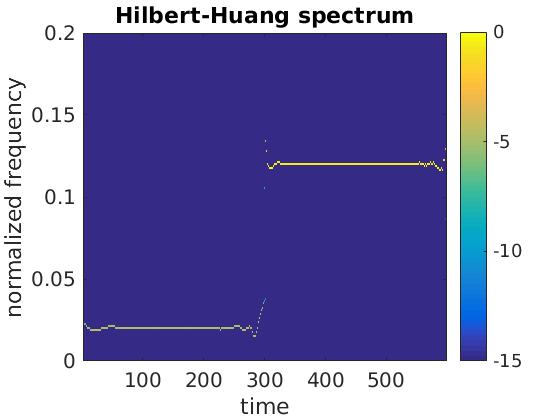}
&
\includegraphics[width = 0.23\textwidth]{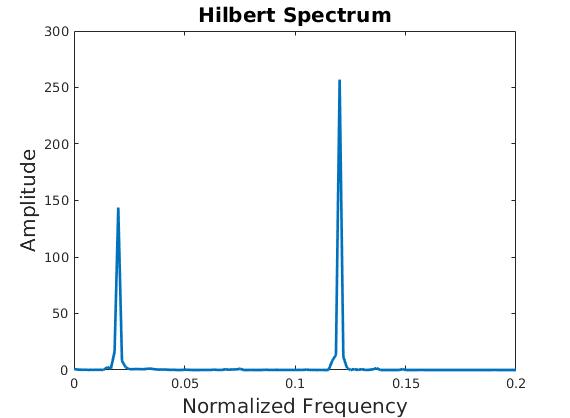}
\end{tabular}
\caption{Hilbert-Huang Spectrum for mode-mixing signal using forward-PDE approach.}
\label{Fig:forward-mode-mixing}

\end{figure}

This experiment addresses the mode mixing separation problem. The signal is built by concatenating two sinusoids with different frequencies, as shown in \Cref{Fig:mode-mixing-signal,Fig:emd-mode-mixing,Fig:forward-mode-mixing}. Unlike classical EMD, the forward-PDE approach can distinguish the two modes and produce a reasonable separation.  As such, it can extract features from mode-mixed signals and obtain better instantaneous frequency details. Classical EMD fails to separate these different modes. 

\subsection{PERFORMANCE MEASURE OF SEPARATION ABILITY}
As shown in \cite{two-freq}, by applying EMD to mixtures of two cosine signals, we can examine the separation capability for different frequency component rations. Consider the following two-component signal, 
$S(x) = cos(2 \pi x) + \alpha cos(2 \pi f x)$,
$\alpha$ is the amplitude ratio and $\alpha \in (10^{-2}, 10^{2})$ and 
$f$ is the frequency ratio, with $f \in (0,1)$. Denote the frequency components as $S_1(x) = cos(2 \pi x)$ and $S_2(x) = \alpha cos(2 \pi f x)$.
Now use the following performance measure for separation capability:
\begin{equation}
PM = \frac{|| IMF_1 - sin(2\pi x) ||_{L^2}}{ || sin(2 \pi x) ||_L{^2} }
\end{equation}

\begin{figure}[h]
\begin{tabular}{ll}
\includegraphics[width = 0.22\textwidth]{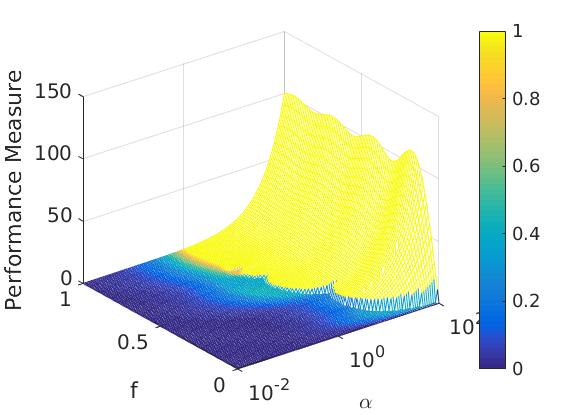}
&
\includegraphics[width = 0.22\textwidth]{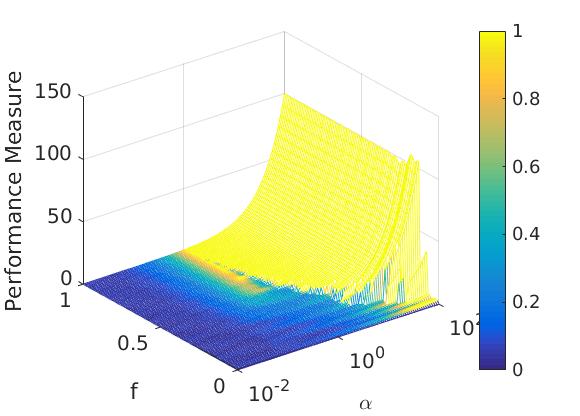}
\end{tabular}
\caption{Left: Performance Measure Regarding $\alpha$ and $f$ for classical EMD. 
Right: Performance Measure Regarding $\alpha$ and $f$ for forward-PDE approach.}
\label{Fig:performance-measure}

\end{figure}

The results are presented in Figure \ref{Fig:performance-measure}.
It should be noted that the performance of our PDE approach is similar to that
of the classical EMD method.

\section{CONCLUSIONS}
This paper presents a forward-PDE modification of the classical EMD algorithm. The mean curve of a signal is obtained by evolving the signal with the heat/diffusion equation and therefore avoids any complicated
methods of extracting local maxima or minima.
 Our approach provides a mathematical interpretation of the EMD algorithm as well as its limitations.  It also performs better on ``mode-mixed'' signals.  Our method also allows the parameters in the PDE to be adjusted according to the properties of the signal being analyzed.

\vfill\pagebreak

\label{sec:refs}


\end{document}